\documentclass[12pt,journal]{article}
\usepackage[utf8]{inputenc}
\usepackage{amsmath}
\usepackage{amsfonts}
\usepackage[margin=1in]{geometry}
\usepackage{latexsym,graphicx,graphics}
\usepackage[nearskip,margin = 0pt]{subfig}
\usepackage{braket,titling}
\usepackage{appendix}
\usepackage{amssymb,epsfig}
\usepackage{amsmath,mdframed,xcolor}
\usepackage{float}
\usepackage{graphicx}
\usepackage{geometry}
\usepackage{tabularx}
\usepackage{graphicx}
\usepackage[makeroom]{cancel}
\usepackage{titlesec}
\usepackage[font={small,it}]{caption}
\usepackage{comment}

\titleformat{\section}
  {\normalfont\fontsize{13}{15}\bfseries}{\thesection}{1em}{}
  \titleformat{\subsection}
  {\normalfont\fontsize{12}{15}\itshape}{\thesubsection}{1em}{}

\def\dul#1{\underline{\underline{#1}}}
\newcommand{\rone}[1]{\boldsymbol{#1}}
\newcommand{\rtwo}[1]{\underline{\underline{\boldsymbol{#1}}}}

\pretitle{% add some rules
  \begin{center}
    \large\bfseries
    }
\preauthor{%
  \begin{center}
    \small \lineskip 0.75em%
    \begin{tabular}[t]{@{}l@{}}%
}
\postauthor{%
    \end{tabular}
    \vskip -.5em
    \par
  \end{center}%
}
\predate{
\small
\itshape
\begin{center}
    }
\postdate{\end{center}}

\title{\vspace{-2.5cm}Enhanced sensitivity deep subwavelength direction-of-arrival sensing using temporal modulation
}
\author{Tamir Zchut, Yarden Mazor}
\date{%
    School Of Electrical Engineering, Tel Aviv University, Tel Aviv, 69978, Israel\\[2ex]%
    \today
}

\begin{document}

\maketitle

\begin{abstract}
Electromagnetic wave interaction with time-varying systems has gained a lot of research interest in recent years. The temporal modulation gives unprecedented control over the response, allowing us to go beyond the state-of-the-art in passive systems. In this work, we use time variation to derive a model for a deep-subwavelength direction-of-arrival (DoA) sensing apparatus with enhanced performance and sensitivity. We formulate the problem, derive an analytical model, and discuss the various physical mechanisms responsible for the enhancement. We show that time modulation enables a new degree of control that can be used to optimize the response for various incident frequencies, allowing for wideband operation. Additionally, we show that incorporating the currents from higher generated harmonics into the sensing scheme allows us to extract more accurate information about the impinging wave.
\end{abstract}

\section{Introduction}
%Direction of arrival sensing is an important application in various fields. This can wait, we can work on this later. The end of the introduction should contain one paragraph and clearly defines the problem we are studying, and how we plan to solve it.
%TZ:The problem is: designing a system that, by tweaking several parameters, will allow us to calculate $\theta$ out of the difference in currents $I_1, I_2$ ??
%YM: Perhaps take a look at the introduction of a a few papers studying DoA detection (Behdad's paper, some others that are cited in our ISF proposal) and try to see how they describe it and learn from them.
% \subsection{Introduction 2}
Detecting a wave's direction of arrival (DoA) has important applications in many fields. Starting from the survival of small species in nature, which rely on understanding the direction a predator is approaching based on the sounds it makes, through aviation and radar systems, and even modern-day imaging technologies such as light-field photography. Usually, DoA detectors rely on the phase differences when the EM signal is recorded by two (or more) adjacent elements, bringing forward a significant limitation - what if the difference is very small, as when the sensing apparatus is small compared to the received signal's characteristic wavelength?

Two main approaches exist in electromagnetic (EM) waves to alleviate this limitation. One approach is based on Bio-mimicking small insects. In large animals, significant ear separation lets information regarding the DoA be extracted from the phase/amplitude difference of the recorded sound. In small insects, this problem is mitigated by direct coupling between the ears. This approach is simple to implement and works very well, mainly around a prescribed angle \cite{BehdadSmallAntennas}. This was followed up by further optimization, including using high-order modes \cite{yi2018subwavelength} and non-foster coupling networks \cite{NonFosterCoup}. 
The second approach is using multiphysics systems. One can couple the EM wave response to a different wave mechanism that operates in similar frequencies but significantly smaller wavelengths, which enhances the recorded signal difference. This is demonstrated for an electro-acoustic system in \cite{HadadPiezo}.

On a parallel route, in the past decade, EM wave interaction with time-varying systems has seen a burst of renewed interest, driven by two parallel processes. First, the ability to implement such systems has seen significant advances, making the theoretical ideas more realistic. Second, the rise of metamaterials has pointed to many intriguing and exotic wave phenomena occurring in materials with extreme parameters. The basic theoretical concepts are introduced in \cite{CassedyPartI1963,Kurth1977}, and numerous applications have been proposed, and among the ones that are more relevant to this work we have the implementation of nonreciprocal elements (gyrators \cite{TimeModGyrator}, circulators and isolators \cite{sounas2013giant,estep2014magnetic}), Nonreciprocal transmission line design \cite{QinCapacitorModulation}, extreme energy accumulation \cite{TretyakovEnergy}, nonreciprocal reflection and transmission \cite{HadadSpaceTime,ShanhuiTimeRef}, and improved antenna matching, Q-factor and bandwidth \cite{AndreaChu,ChuParametric,ParametricChuAluExperiment,HadadIndirect} (many other applications are summarized in \cite{CalozSpacetime,MicrowaveNonreciprocity}). Of specific relevance are the control of Mie scattering properties, shown in \cite{TretyakovTimeModScatterer}.  Non-periodic variation and switching also play an important role, demonstrated for broadband matching \cite{HadadBodeFano}, mathced filter design \cite{SilbigerHadad}, engineered time-reversal spectrum \cite{GaliffiYinAlu}, and many more.

In this work, we will examine how we can enhance the sensitivity of DoA detection in deeply subwavelength systems using a different approach - by incorporating temporal modulation to enhance and tailor the physical interactions between the system elements. Since time-modulated systems naturally have many parameters at play, and rich content of physical processes occurring, we will focus on unveiling the physical mechanisms at play that enable sensitivity enhancement. 

Several works have already started examining this avenue of time-modulated DoA systems. First, in \cite{ModDOA} and subsequent works, periodic switching of the received signal from a DoA sensing dimer is used to enable angle estimation through the ratio of power content in various higher harmonics. However, since only the received signal is switched, physical interaction between the elements does not benefit from these higher harmonics. Moreover, the coupling between the elements, which will be pivotal in this work, does not play a significant role. This makes this strategy viable mostly for $d>\lambda/2$, and not the deeply subwavelength scenario we target. In \cite{BillotiDOA}, a time-modulated metasurface was proposed for DoA detection, leveraging the deviations in the diffraction/reflection angles to do the estimation. Again, this work utilizes a system whose size is on the order of $\lambda$.
%Recently, non-periodic variation of the media based on abrupt space-time discontinuities \cite{deck2019uniform} has also been proposed for broadband matching \cite{HadadBodeFano} and anti-reflection coatings \cite{NaderAntiRef}. One key aspect of these applications is that adding time-variation allows us to escape the constraints of fundamental physical bounds imposed by the linearity and time-invariance of the wave system, e.g., The Bode-Fano bound \cite{BodeBook} and small antenna Chu-Harrington limit \cite{ChuLimit,harrington1960effect}.

\section{Formulation}
\begin{figure}[h]
\begin{center}
\noindent
  \includegraphics[width=4.7in]{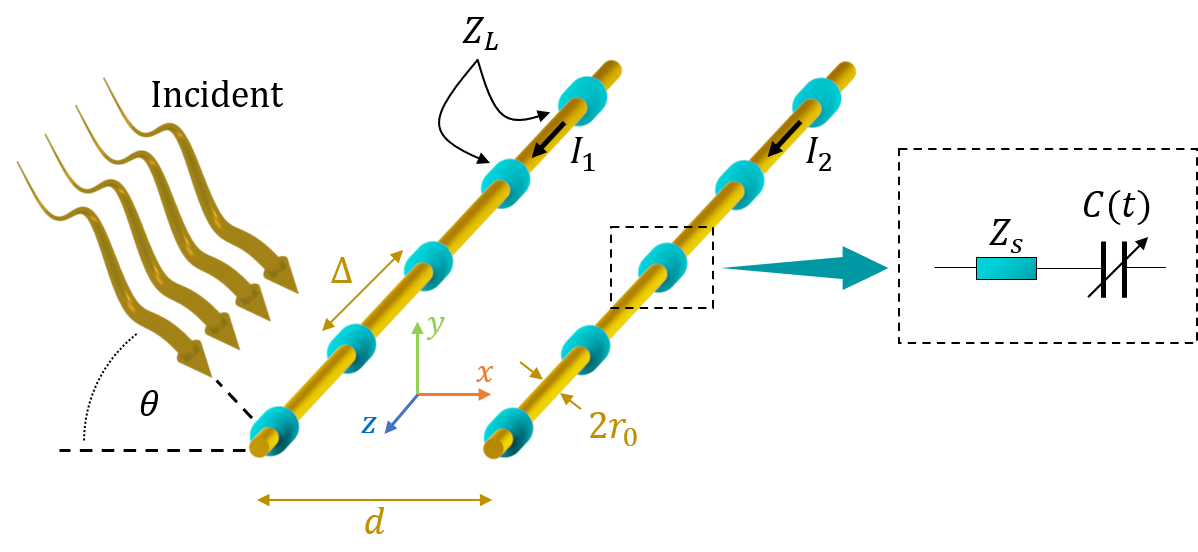}
  \caption{Dimer model: two infinite wires, with periodic loading $Z_L=Z_S+Z_C$, where $Z_S$ is a static impedance, and $Z_C$ is the impedance of a time-modulated capacitor.}
  \label{fig:basicDimer}
\end{center}
\end{figure}

%How is our model system constructed?
%TZ: what does it mean?
%YM: Just a description of our system in a clear way. Dimer, loading, period, etc.
Naturally, time-modulated wave systems possess many degrees of freedom. On top of the "regular" time-invariant parameters, we also have the modulation frequency, depth, waveform, and the ensuing coupling to higher and lower frequencies. Due to that, the analysis becomes complicated quite rapidly with the addition of components. Since we aim to focus on the physical mechanisms that enhance the DoA sensitivity, we employ a simple model - a 2D dimer. To explore our concept, we use a dimer composed of two infinite wires made of a perfect electric conductor (PEC), as shown in Fig. \ref{fig:basicDimer}. The incident wave angle $\theta$  corresponds to a phase difference of $kd\cos(\theta)\ll 1$. The wires are periodically loaded with $Z_p=Z_s+Z_C$, where $Z_s$ is a static, non-modulated impedance, and $Z_C$ is a time-modulated capacitor $C(t)$.
%Modelling the response of a single wire.
The wires are modeled using their susceptibility $\alpha$, which determines the induced current $I(\omega)$ (without modulation) on each wire via
\begin{equation}
    I(\omega)=\alpha(\omega) E^{loc}_{tan}(\omega),
\end{equation}
Where $E^{loc}_{tan}(\omega)$ is the tangential component of the local field, which contains both the incident wave, and the field scattered by the other wire, but in the absence of the examined wire itself. For infinite PEC wires, the susceptibility is \cite{hadad2019spacetime,tretyakovBook}
\begin{equation}
    \alpha^{-1}(\omega)=\alpha_0^{-1}(\omega)+\frac{Z_L}{\Delta},\qquad \alpha_0^{-1}(\omega)=\frac{\eta k}{4}H_0^{(2)}(kr_0)
\end{equation}
Where $\alpha_0$ is the susceptibility of the unloaded wires, $Z_L$ is the $\Delta$-periodic loading, $\eta$ is the medium impedance, $k$ is the free-space wavenumber, and $H_0^{(2)}$ is zero-order Hankel function of 2nd kind. To clarify the notation, later we will use also $\gamma=\alpha^{-1},\gamma_0=\alpha_0^{-1}$, and for simplicity, we focus on TE incidence (therefore $\hat{z}$ electric field).
% \begin{equation}
%     E(t)=\int_{-\infty}^{\infty}\gamma(t')I(t-t')dt' \label{Singlewirereaction}
% \end{equation}
%Modelling the response of a dimer. 
For every frequency $\omega$, the current on the wires generates scattered fields via $E^{scattered}_{tan}(\rho)=G(\rho,\omega)I$, where $G(\rho,\omega)$ is the 2D Green's function of the electric field in free-space, i.e.  $G(\rho,\omega)=-\frac{\eta k(\omega)}{4}H_0^{(2)}\left(k(\omega)\rho\right)$.\\
Using both $G$ and $\alpha$, we can express the total reaction of a non-modulated dimer are
\begin{align}
    I_2=\frac{\alpha_2 E_2^{inc}+G(d)\alpha_1\alpha_2 E_1^{inc}}{1-G^2(d)\alpha_1\alpha_2}, \;I_1=\alpha_1\left(E_1^{inc}+G(d)I_2\right).\label{currents}\smallskip
\end{align}
Now, let us incorporate the effects of periodic temporal modulation into the system. When the capacitor $C(t)$ is periodically modulated, $ C(t)^{-1}=C_0^{-1}[1+m\cos(\omega_m t+\varphi_m)]$, we can represent every physical quantity as a sum of all the possible harmonics
\begin{equation}
    X(t)=\sum^{\infty}_{n=-\infty} \tilde{X}_ne^{j(\omega+n\omega_m)t}+c.c.=\sum^{\infty}_{n=-\infty} \tilde{X}_ne^{j\omega_nt}+c.c.
    \label{eq:expansion}
\end{equation}
Where X can be the current $I$ or the electrical field $E$ or any other relevant quantity, $\omega_n=\omega+n\omega_m$, and c.c. means complex conjugate.
%How do we incorporate time modulation into the response of a single wire.
Since both the current $I$ and the electric field $E_z$ are excited in multiple frequencies (As expected in our temporally modulated system), the amplitudes of the spectral components can be ordered in a column vector $[\tilde{I}]$, and $[\tilde{E}]$. Using these, and following \cite{Kurth1977,hadad2019spacetime}, we can write the time-modulated single wire response:
\begin{equation}
    [\tilde{E}] = \rtwo \Gamma [\tilde{I}]\leftrightarrow
    \begin{bmatrix}
    \vdots \\
    \tilde{E}_{-1} \\
    \tilde{E}_{0} \\
    \tilde{E}_{1} \\
    \vdots
    \end{bmatrix}=
    \begin{bmatrix}
    \ddots & \vdots & \vdots & \vdots & 0\\
    \hdots &  \gamma_0(\omega_{-1})+\frac{1}{j\omega_{-1}C_0\Delta} & \frac{M}{j\omega_{0}C_0\Delta} & 0 & \hdots \\
    \hdots &  \frac{M^*}{j\omega_{-1}C_0\Delta} & \gamma_0(\omega_{0})+\frac{1}{j\omega_{0}C_0\Delta} & \frac{M}{j\omega_{1}C_0\Delta} & \hdots \\
    \hdots &  0 & \frac{M^*}{j\omega_{0}C_0\Delta} & \gamma_0(\omega_{1})+\frac{1}{j\omega_{1}C_0\Delta} & \hdots \\
    0 & \vdots & \vdots & \vdots & \ddots 
    \end{bmatrix}
    \begin{bmatrix}
    \vdots \\
    \tilde{I}_{-1} \\
    \tilde{I}_{0} \\
    \tilde{I}_{1} \\
    \vdots
    \end{bmatrix},\label{eq:gammaMatrix}
\end{equation}
Where $M=me^{j\varphi_m}$.
Since the current is excited in several frequencies, each of these will generate a scattered field at it's specific frequency. Since $G$ is a function of the frequency $\omega$, it can be formulated using a diagonal matrix, $\rtwo G$ 
\begin{equation}\label{eq:gMatrix}
    \rtwo G (\rho)=diag[...,G(\rho,\omega_{-1}),G(\rho,\omega_0),G(\rho,\omega_1),...]
\end{equation}
%Incorporating time modulation into the dimer response.
Then, for the modulated dimer, the currents can be solved from the following matrix equation
\begin{equation}
    \rtwo \Gamma_1[\tilde{I}_1]-\rtwo G [\tilde{I}_2]=[\tilde{E_1^i}]\;,\;\rtwo \Gamma_2[\tilde{I}_2]-\rtwo G[\tilde{I}_1]=[\tilde{E_2^i}]
    \label{eq:dimerMod}
\end{equation}
%Where $[\tilde{I}]$ and $[\tilde{E}]$ are vectors of the expansion coefficients in Eq.  \ref{eq:expansion}, and $\underline{\underline{\Gamma}}$ includes the wires response, and the coupling between harmonics due to the modulation.

%The formulation we use is based on  representing the response of each wire using it's polarizability $\alpha$.
\subsection{Performance metrics}

%Defining S
In order to compare the performance and possible benefits of our approach, both with respect to different modulation scenarios and other enhancement schemes, we use the sensitivity $S$, defined as
\begin{equation}
    S(\theta)=\left|1-\frac{I_1(\theta)}{I_2(\theta)}\right|^2
\end{equation}
which essentially quantifies the relative variation of $I_1,I_2$ with respect to the incident angle. To accurately extract $\theta$ from $I_1,I_2$, $S$ should vary with large amplitude against $\theta$. 
%Defining Smax

In addition to the sensitivity, we would also like to define a more simple metric, that will allow us to capture the properties of $S(\theta)$ using a single scalar. This will allow us to characterize the sensitivity curve against various system parameters (wire loading, temporal modulation parameters). To that end, we will use the maximum sensitivity, $S_{max}$:
\begin{equation}
    S_{max}=\max_{\theta\ }\{S(\theta)\}
    \label{eq:Smax_def}
\end{equation}
In the modulated system, $S_{max}$ can be calculated for the different up- and down-converted harmonics (which will prove very beneficial), and therefore we will often use $S_{max}^n$ to indicate that it was calculated for the n'th harmonic, where $n=0$ is the fundamental, incident frequency. $S_{max}^{NM}$ (or any other occurance of $NM$) will indicate the non-modulated case.
% Following the discussion for $S$, we aim for this number to be large, to indicate strong differences between currents vs. $\theta$.

%Explaining why are these good performance metrics

\section{Results and Discussion}
\subsection{Frequency conversion effects}\label{sec:frequencyConv}
\begin{figure}[H]
\centering
\includegraphics[width=10cm]{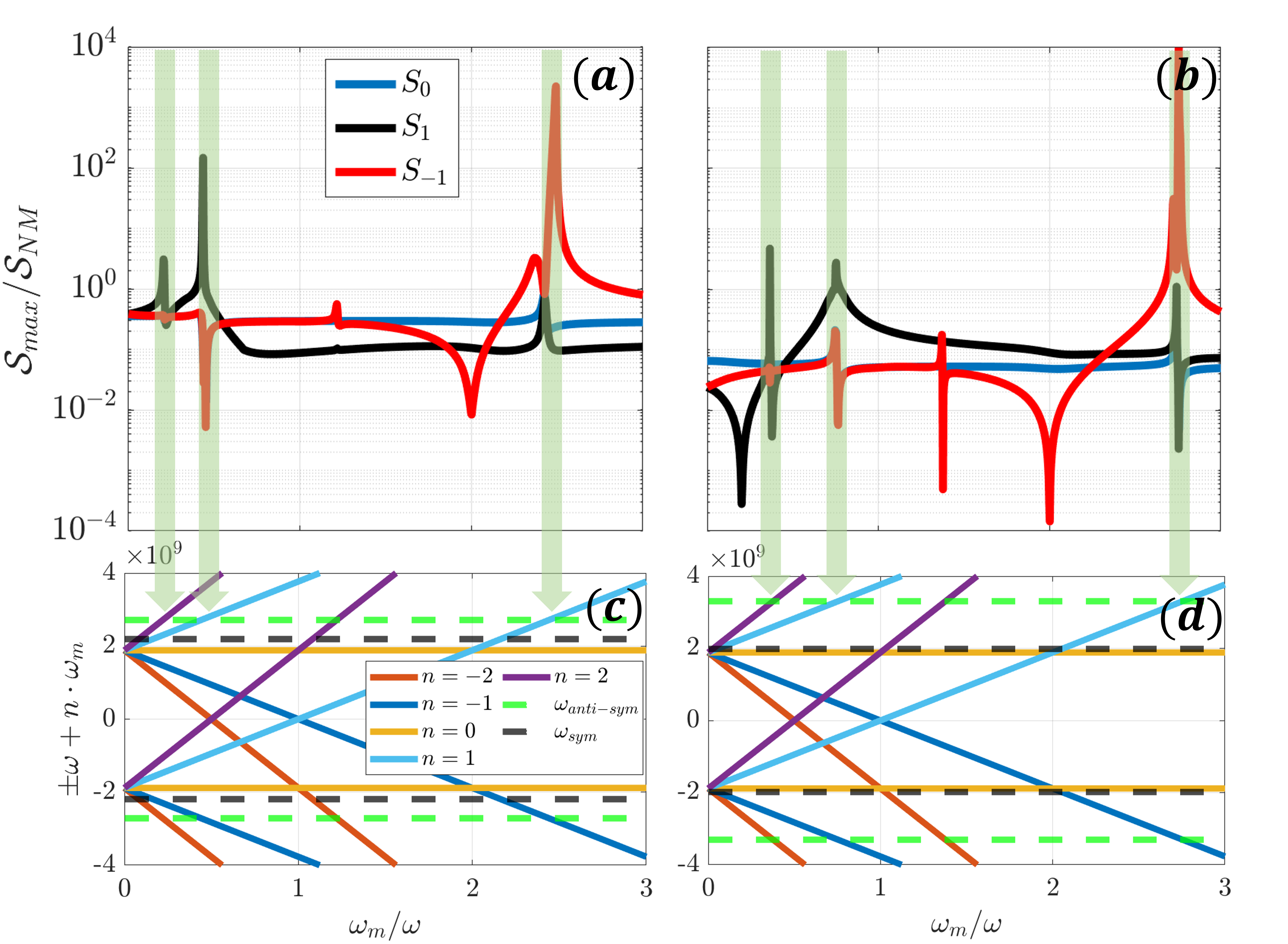}
\caption{(a) maximum sensitivity as a function of $\omega_m/\omega$ for harmonies $-1,0,1$, and $d=5cm$ (b) same, for $d=1cm$ (c) frequency conversion map for $\pm\omega+n\cdot\omega$, with $-2\leq n\leq 2$, and the symmetric and anti-symmetric resonance frequencies. The green arrows indicate connection between peaks in the sensitivity and intersection in the resonance map. (d) Same, for $d=1cm$.}
\label{fig:Resonancemap}
\end{figure}

Various physical mechanisms play a role in enhancing the sensitivity of the dimer response to the DoA. To discuss these systematically, let us start by looking at the $S^n_{max}/S_{max}^{NM}$ vs. the modulation frequency. Figure \ref{fig:Resonancemap} shows $S_{max}^{\{-1,0,1\}}$ for the following parameters: The radius of the wire is $r_0=0.3mm$, the distance between the wires is $d=5cm$, and the base frequency is $f_0=300MHz$. The period of loading is $\Delta=1cm$, with the load consisting of a periodic resistor $R_L=0.3m\Omega$ , and a capacitance $C^{-1}(t)=C_0^{-1}(1+m\cos(\omega_m t))$, where $C_0=13pF$ and $m=0.2$.
In the fundamental frequency, we see that mostly $S^{max}$ is similar to the non-modulated case (yielding a normalized value close to 1), except for a small region around $\omega_m\approx  2.4\omega$ which exhibits a moderate improvement, $S^0_{max}/ S_{max}^{NM}\approx 2.7$. However, when looking at the higher harmonics, we see that the sensitivity becomes much higher at certain values of $\omega_m$ ($\omega_m\approx 0.21\omega,0.42\omega,2.4\omega$). The effect behind this is composed of the interaction of two mechanisms: the eigenmodes of the dimer and frequency conversion induced by the temporal modulation.
In the unmodulated dimer, each wire, incorporated with capacitive loading, is resonant (since the pristine wire response is inductive). When two such resonant wires are placed next to each other to form a dimer, two resonance frequencies exist. Each resonant frequency is characteristic of a specific dimer mode - a symmetric and an anti-symmetric one that form a basis for any current combination that may be excited in the wires. We denote the resonance frequencies of each of these modes $\omega_{sym},\omega_{anti-sym}$ respectively. How strongly each dimer mode contributes to the total currents is a function of the excitation frequency and the exciting field distribution. Since we want to extract information regarding the DoA from the differences between the currents, we would like to "boost" the content of the anti-symmetric mode as much as possible and how steeply this content varies as a function of the incidence angle (considering that for certain values of $\theta$, namely $\pi/2$ and $3\pi/2$, there is only the symmetric mode, regardless of other parameters, due to symmetry considerations). 

When considering a deep subwavelength dimer, we expect the content of the symmetric mode to be dominant, resulting from the slight difference in exciting field between the dimer wires. This makes the current differences between the wires vary only slightly when changing $\theta$, making it harder to extract information about the DoA.  
%Due to this symmetric content being prevalent, very small current differences are induced when  changing twe which is 2 identical wires, we can see it is symmetrical, meaning that when exciting the system, we would expect the symmetric mode to be stronger, and the anti-symmetric mode to be very weak, since our system is symmetric. Hence, we expect the sensitivity to be much smaller for $S^{NM}_{max}$.\\

When adding periodic temporal modulation, another mechanism behind the sensitivity enhancement comes into play - frequency conversion. When the capacitor is allowed to vary with time, we convert the fields in the basic problem to higher (and possibly lower) frequencies, i.e. $\omega_n=\omega_0+n\omega_m$. When $\omega_m$ is chosen such that one of the $\omega_n$ frequencies coincides with $\omega_{anti-sym}$ we can obtain a strongly enhanced anti-symmetric mode content in the corresponding n'th harmonic, which is also steeply dependant on the incident angle $\theta$. This results in a enhanced $S^n_{max}$ for that corresponding harmonic, as seen in figure \ref{fig:Resonancemap}(a,b) for dimer separation $d=5cm,1cm$ respectively.

%What physical mechanisms are in play? 

%Frequency conversion to the anti-symmetric resonance - 
%show plot of optimal modulation frequencies vs. Smax. 
%Show a chart of the harmonics explaining this plot. Show the possible tunability of the optimal angle.
Using Fig. \ref{fig:Resonancemap}(c,d), we can see the correlation between these two mechanisms, which enhance the sensitivity. 
The continuous lines represent different $\omega_n$s as a function of $\omega_m$, and the dashed lines show the resonance frequency of the symmetric and anti-symmetric modes in the unmodulated dimer. When a continuous line intersects the green dashed line, $\omega_n=\omega_{anti-sym}$ is satisfied for that n'th harmonic, the $\omega_m$ values that yield this intersection are roughly the modulation frequencies where we experience a sharp increase in the sensitivity. 
Thus, by choosing the dimer parameters, we can alter $\omega_{anti-sym}$ and $\omega_m$, and tailor the frequency response we need, to sense the expected $\theta$s better for the required range of operating frequencies. It is essential to add here that while this proves good physical intuition, quantitatively, there are minor deviations in the predicted value of $\omega_m$ based on this intuition and the actual value in which the enhancement is most prominent. This is due to the fact that we make use of the resonance frequencies in the \textit{unmodulated system}. These frequencies also experience a shift under modulation since the effective impedance of the wires changes. 

% \begin{comment}
% As can be seen in fig.\ref{fig:contourCandL} by changing the load to $Z_L=R_L+Z_{C,modulated}+\frac{1}{j\omega_n C}$ in fig. \ref{fig:contourC} or to $Z_L=R_L+Z_{C,modulated}+ j\omega_n L$, we can add an additional degree of freedom that controls the optimal modulation frequency and the degree of enhancement. The intuition we have previously acquired remains useful here too: in white, we see the modulation frequency corresponding to conversion to the symmetric resonance, while in red we see the value for the anti-symmetric resonance. It can be seen the peaks in the sensitivity follow exactly the anti-symmetric path.
% One could also incorporate resonant impedances into the load, to achieve more enhancement peaks due to additional anti-symmetric resonances.    
% \end{comment}
% This effect can also be altered and enhanced, by modifying the periodic loading with additional passive elements. 
Next, we would like to explore additional venues to tune the performance of our dimer. The natural way to do that, is by tailoring the periodic loading $Z_L$ using additional passive elements such as inductors, capacitors, or a combination of the two.

To simplify the calculation we add the components in series. When adding either an inductance $L$ or capacitance $C$, the load impedance $Z_L$ in the n'th harmonic becomes $Z_L(\omega_n)=R_l+\frac{1}{j\omega_n C_m}+j\omega_n L$ or $Z_C(\omega_n)=R_l+\frac{1}{j\omega_n C_m}+\frac{1}{j\omega_n C}$ respectively. Figure 3 presents the sensitivity dynamics as a function of the values of $L,C$. In panels (a,b,d,e) we see the sensitivity of both the base harmony $S_0$ and the first harmony $S_1$, as a function of the added passive element ($L$ for (a,d) or $C$ for (b,e)) and of the normalized modulation frequency $\frac{\omega_m}{\omega}$. In all these panels, we see that as the passive element value, $L$ or $C$, changes, the modulation frequency of the peak $\omega_m^{peak}$ changes as well. %\com{Please describe better. What increases? what decreases?}
%, the peak of the maximum sensitivity is transitioning to lower values of $\omega_m$ as the value of $L$ or $C$ increases, for both $S_1$ and $S_0$. 
% Hence, adding inductance or capacitance can modify the sensitivity of the system, allowing us to better tune the modulation frequency to achieve maximum sensitivity in the operating frequency. \\
%\com{How does this relate to the red and yellow lines that we see too?}
In these panels, the red line shows the location of the anti-symmetric resonance frequency for the non-modulated circuit, and the yellow lines show the location of the symmetric resonance frequency. As portrayed before, the sensitivity peak modulation frequency moves together with the conversion frequency corresponding to the anti-symmetric mode. These panels show that this concept can be generalized, and the loading can be used to tailor the response to our needs. 

\begin{figure}[H]
\centering
\includegraphics[width=\textwidth]{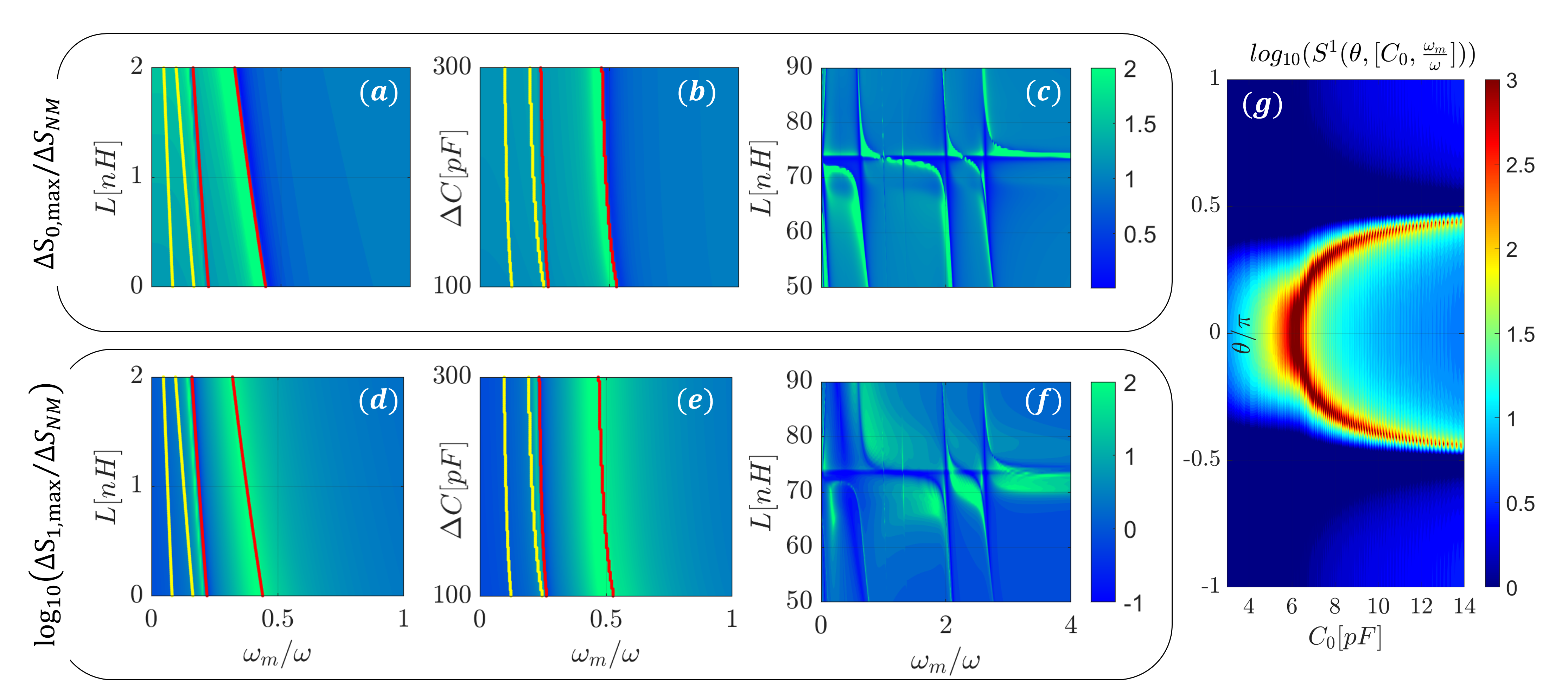}
\caption{(a,b,c): Normalized maximum sensitivity in the fundamental harmonic, as function of $\omega_m/\omega$, and of the value of inductance/capacitance added to $Z_L$. (d,e,f):Normalized sensitivity of the 1st harmonic, as function of $\omega_m/\omega$, and of the value of inductance/capacitance, on a logarithmic scale. The red lines are the evaluated $\omega_m$ required for the first harmonic to coincide with the anti-symmetric resonance frequency of the non-modulated circuit (yellow for the symmetric resonance). for (c,f): $C=3pF$. (g): $S^1(\theta)$ for varying values of $C_0$.}
\label{fig:contourCandL}
\end{figure}

When combining inductance and capacitance, we introduce more resonance frequencies into the non-modulated system, which provide additional sensitivity peaks when coinciding with converted harmonics. For example, when using $Z_L(\omega_n)=R_l+\frac{1}{j\omega_n C_m +\frac{1}{j\omega_n L +\frac{1}{j\omega_n C}}}$ we see in Fig. \ref{fig:contourCandL}(c,f) how additional peaks for different $\omega_m$ values are added to the system, which can be used to design a multi-frequency sensing dimer.
%Adding various loads to tailor the response (capacitive, resonant)
In fig. \ref{fig:contourCandL} (g), we see how a change to the modulation capacitor controls the specific $\theta$ around which the sensitivity changes most drastically. For each value of $C_0$ the optimal $\omega_m$ is chosen, and $S^1$ is calcualted.
%Frequency conversion to high harmonics where $d\ll\lambda$ is no longer satisfied.
\subsection{Energy balance and conversion efficiency}
% \begin{comment}
%     \com{add here the discussion about energy balance vs. the modulation frequency, and also discuss the conversion efficiency to higher harmonics. Perhaps also show how the scattered energy in n=1 and n=-1 is in relation to the scattered energy for the fundamental frequency (n=0).}
% \end{comment}
Since our system is time-modulated it is not, in general, passive. Here, we would like to examine the balance between the incident power used to excite the currents $P_{inc}$ and the radiated power $P_{rad}$ to better understand the energy dynamics. 
% \begin{comment}
% \com{The derivation should be much more concise here. It's written in a way that's more suitable for the Thesis document than for a paper. The full derivation can always be added as additional material in an appendix. I would start by describing equation 11 as the balance between the work done to drive current in the wires, and the radiated energy (in a lossless system). Then, say that given that you calculate the wire currents in each frequency, you can calculate the radiated power (say $P_{radiated}$) using equation 23, and the power delivered to the wires by equation 15. Then, explain that when no modulation exists, these quantities are in equilibrium, but due to the modulation this balance is now violated.}
% \end{comment}
Assuming that no material losses are present, the only way energy can exit the system is through radiation. In reality, there are some losses to the wires and the system elements, but the dissipated power is negligible in comparison with the radiated power. If we define a cylindrical envelope $\mathcal{S}$ with radius $R\rightarrow\infty$, the radiated power given by (Appendix)
% Using Poynting Theorm, removing zeroed elements, we get the balance between the work done to drive current in the wires on the LHS, and the radiated energy in a lossless system on the RHS:
% \begin{equation} \label{eq:PoyntingTheorem_Radiated}
%     \frac{1}{2}\Re \left\{ \iint (\rone{E} \times \rone{H}^*) \cdot \rone{dS} \right\}=\frac{1}{2}\Re \left\{ \iiint \rone{E}\cdot\rone{J}^* dV \right\}
% \end{equation}
\begin{equation} \label{eq:LHSdimernomod}
    P_{scat}(\omega)=\frac{\eta^2 k}{4}\left(\left[|I_1|^2+|I_2|^2\right]+2J_0(kd)\Re\left\{I_1\cdot I_2^*\right\}\right)
\end{equation}
where $J_0(z)$ is 0th order Bessel function. On the other hand, the power extracted from the incident field, $P_{inc}(\omega)$ is given by 
\begin{comment}
In order to calculate the RHS, we start with:
\begin{equation}
     RHS=\frac{1}{2}\Re\left\{\bar{E}^{inc}_{wire}\bar{I}^*_{wire}\right\}+\frac{1}{2}\Re\left\{\bar{E}^{inc}_{other-wire}\bar{I}^*_{other-wire}\right\}
\end{equation}
Where:
\begin{equation}
    \bar{E}_2^{inc}=\bar{E}_1^{inc}\cdot e^{jkd\cos(\theta)}=\begin{bmatrix}
        \dots,0,1,0,\dots
    \end{bmatrix}\cdot e^{jkd\cos(\theta)}
\end{equation}
With the $1$ is at frequency $\omega_0$, since the incident wave have a frequency of $\omega_0$.\\
The current, for each wire:
\begin{equation}
\bar{I}_{wire}=\dul{\alpha}\bar{E}_{wire}=\dul{\alpha}\left(\bar{E}^{inc}+\dul{G}(d)\bar{I}_{other-wire}\right)
\end{equation}
With $\dul{\alpha}=\dul{\Gamma}^{-1}$, where $\dul{\Gamma},\dul{G}$ are defined in eq. \ref{eq:gammaMatrix} and eq. \ref{eq:gMatrix}, respectively.\\
Since the problem is symmetric, we can deduce it to both wires:
\end{comment}
% Given the currents on the wires are calculated in every frequency using eq. \ref{eq:dimerMod}, we can calculate the power delivered to the wires, $P_{inc}$:
\begin{equation}
    P_{inc}(\omega)=\frac{1}{2}\Re\left\{\alpha^*(|\bar{E}^{inc}_1|^2+\bar{E}^{inc}_1 G^*(d)\bar{I}^*_{2})\right\}+\frac{1}{2}\Re\left\{\alpha^*(|\bar{E}^{inc}_2|^2+\bar{E}^{inc}_2 G^*(d)\bar{I}^*_{1})\right\}
\end{equation}
When the system is not modulated and lossless, these quantities will be in equilibrium. When time-modulation is introduced, this equilibrium is violated, since the modulation can potentially provide additional power to the system (or act as a sink and extract power). In this case, there are additional frequencies in which current is generated, and therefore power is radiated, rendering the total scattered power as a sum over all possible frequencies $P_{scat,tot}=\sum_n P_{scat}(\omega_n)$. 
% \begin{figure}[H]
% \centering
% \includegraphics[width=\textwidth]{En_sc_tot_divide_En_inc(wm,maxmintheta)_mod_dimer.png}
% \caption{Sum of scattered energy, normalized by the incident energy, for maximum and minimum gain as a function of $\theta$.}
% \label{fig:Energynormscattered}
% \end{figure}
% \begin{comment}    
% In fig. \ref{fig:Energynormscattered}, the normalized sum of energy harmonics is displayed, with regards to the maximum and minimum gain we can get as a function of $\theta$. By looking at the different peaks in this graph, we can divide them into two types: peak of energy and peak of gain. In the peak of energy, the normalized sum is getting bigger, but there is no difference in gain, for example the peak around $\frac{\omega_m}{\omega_0}=2.1$. On the other hand, in the peak of gain, the total energy can be small, but there is a big difference in gain between the max and min gain, for example the peak around $\frac{\omega_m}{\omega_0}=2.5$.
% When comparing the frequencies of the peaks to the resonances map on fig. \ref{fig:Resonancemap}(c), we see that frequencies of energy peaks are the intersection frequency between an harmonic and the the symmetric resonance frequency. On the other hand, the frequencies of gain peaks are the intersection between an harmonic and the anti-symmetric resonance frequency. Hence, we can deduce that in the points of intersection between the harmonics and the anti-symmetric resonance frequency we get an amplification to the expected gain of energy.\tamir{I need to better explain it.. and add more deductions}
% \end{comment}
%To 
\begin{figure}[H]
\centering
\includegraphics[width=11cm]{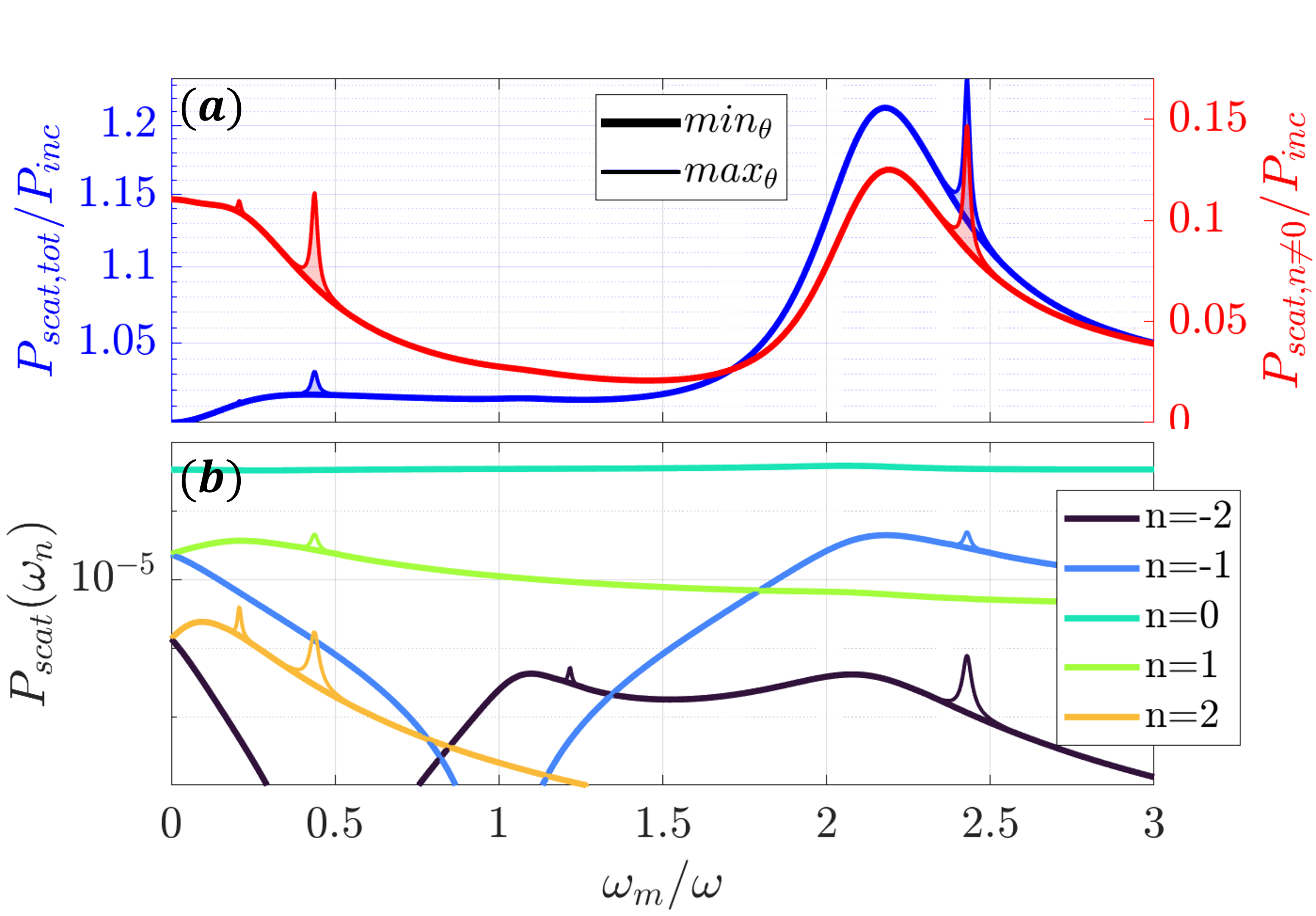}
\caption{(a) Normalized sum of the total scattered energy (in all harmonics) (blue), and normalized sum of scattered energy in the up- and down-converted harmonics (excluding the fundamental $n=0$) of the scattered energy (red).(b) scattered energy of 5 harmonies $-2<n<2$. Both are functions of the normalized modulation frequency $\frac{\omega_m}{\omega_0}$, and show the difference between the minimal energy as a function of $\theta$ (thick line), and the maximum energy as a function of $\theta$ (thin line).}
\label{fig:EnergymodNorm}
\end{figure}
In Fig. \ref{fig:EnergymodNorm}(a), we examine the total power gain (blue) and the up- and down-converted scattered power as a function of the modulation frequency $\omega_m$. For each choice of $\omega_m$ the obtained scattered power is also a function of the incidence angle $\theta$, and since we would like to examine these dynamic, we represent each quantity by two different lines: a thick line, which depicts the minimal power as a function of $\theta$, $\min_\theta \{P_{scat}\}$, and a thin line, which depicts the maximal energy as a function of $\theta$, $\max_\theta \{P_{scat}\}$. Across all examined modulation frequencies a small amount of total gain is provided to the system, since the blue line in \ref{fig:EnergymodNorm}(a) is always $>1$. Additionally, in certain values of $\omega_m$ we see there is a significant difference between the maximum and minimum for different DoA $\theta$.
% This augments the effect introduced before, demonstrated in the resonance map in \ref{fig:Resonancemap}(c) We see that the modulation frequencies of the peaks, are the same intersection frequencies in fig. \ref{fig:Resonancemap}(c). 
This is a complementary mechanism - when the modulation frequency up- or down-converts to the anti-symmetric dimer mode, the gain provided to the system has a significant variation as a function of $\theta$, playing a role in the resulting enhanced sensitivity to the DoA. Figure \ref{fig:EnergymodNorm}(b) helps us confirm this picture. We see that the total scattered energy in each harmonic experiences a noticable dependance on $\theta$ (manifesting as a gap between the thick and thin lines) when the conversion corresponds to the anti-symemtric dimer mode.
%\com{Tamir, what about the red line?}
When examining the normalized sum of the converted harmonics (all $n\neq 0)$ (red), we see that for modulation frequencies corresponding to conversion to dimer resonances the scattered power is $~3$ times larger, which indicates the much higher conversion effciency in this regime. 
% a certain effect is emphasized around the higher modulation frequencies, where two peaks appear around $\omega_m\approx 2.3 \omega_0$. In these modulation frequencies, the energy of the up- and down-converted harmonies becomes bigger. This is due to both higher conversion of energy from the base harmony to higher harmonies around those modulation frequencies, and energy provided by the modulated element.

% These frequencies enhance the anti-symmetric mode content, thus enhancing the difference between the current. They also cause an excessive translation of energy from the basic harmony to other harmonies, for certain values of DoA $\theta$. In these points of $(\omega_m,\theta)$, the currents are in the resonance frequency of the anti-symmetric mode, and have a big enough difference to have a significant anti-symmetric part to be amplified. Hence there is an amplification to the currents, which translated into bigger scattered energy.
In fig. \ref{fig:EnergymodNorm}(a)(red), we notice two types of peaks in sensitivity - peaks for which the amplitude varies as a function of $\theta$ (manifesting as a significant gap between maximum and minimum values), and peaks which do not have this "spreading." Another characteristic that is different between these is the width of the peaks. The peaks associated with the anti-symmetric mode are much narrower than the symmetric one. This is due to the fact that the anti-symmetric mode has a significantly higher quality factor (Q factor), a consequence of the fact that the primary "loss" mechanism here is radiation, and the anti-symmetric mode radiates much less efficiently. This can be seen using the scattered energy equation, eq. \ref{eq:LHSdimernomod} - for given current magnitude on each wire, when the currents are out of phase ($I_1=-I_2$) we have $I_1\cdot I_2^* <0$, while for the symmetric mode $I_1\cdot I_2^*>0$. Since $J_0(kd)\approx 1$, the expected radiated power will be much higher for the symmetric mode than for the anti-symmetric mode.

\subsection{Fast modulation}
\begin{figure}[H]
\centering
\includegraphics[width=9cm]{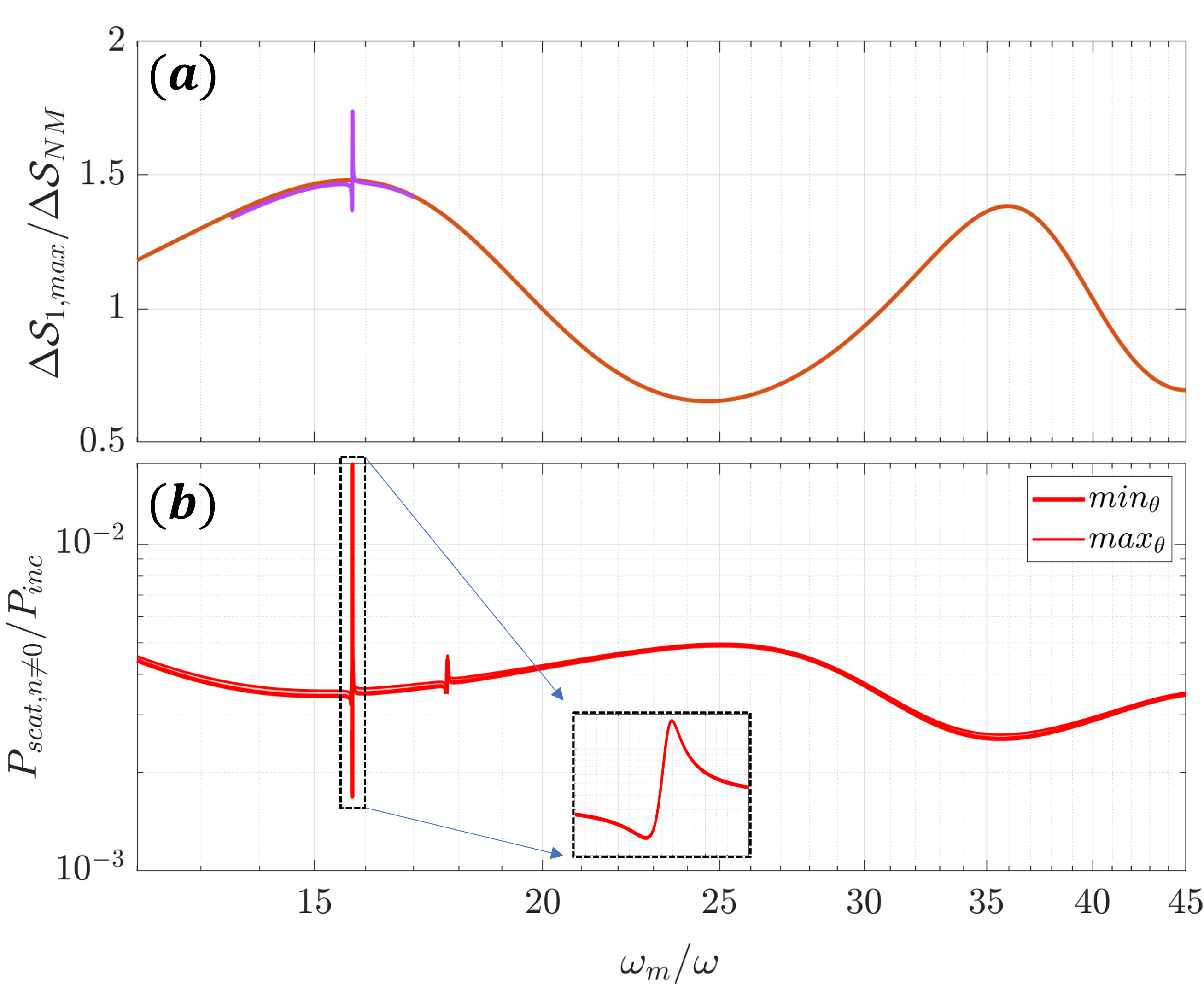}
\caption{Top: $S^1_{max}$ as a function of $\frac{\omega_m}{\omega}$, with $10<\frac{\omega_m}{\omega}<45$. At $\frac{\omega_m}{\omega}>20$, we see ripples in the sensitivity. Bottom: normalized energy as function of $\frac{\omega_m}{\omega}$.}
\label{fig:contourLextended}
\end{figure}

Another possible enhancement mechanism is revealed when examining higher modulation frequencies. While it might be challenging to achieve fast modulations (depending on the incident wave frequency and other system paramters), this physical mechanism exists and may be of use. When examining $\Delta\mathcal{S}_1$ for higher values of $\omega_m$, we notice "ripples" that start occuring, when plotting $\Delta\mathcal{S}_1(\omega_m/\omega)$, as seen in figure \ref{fig:contourLextended}(a), in red. 
%interesting effect that can be harnessed to increase $S_{max}$ is a ripple effect that can be seen in higher frequencies:
This effect happens because the dimer is no longer deep-subwavelength at higher harmonics when the modulation is fast enough. Therefore, the interaction between the wires through the surrounding medium contributes a non-negligible phase difference which causes the exciting field (incident + interaction) to couple better with the antisymmetric dimer mode, enhancing the sensitivity. Since this enhancement is non-resonant, it is milder than we saw previously. However, it can be incorporated with resonance by loading the wires with a resonant impedance in these frequencies, as shown in magenta in figure \ref{fig:contourLextended}. The basic parameters are the same as in section \ref{sec:frequencyConv}, and the additional resonant loading consists of $L_1\approx 5.24nH,C_1\approx 0.195pF$ connected similarly to the results shown in Fig. \ref{fig:contourCandL}(c,f). These are also demonstrated in the scattered power in Fig. \ref{fig:contourLextended}(b). if we examine the enhancement around $\omega_m\approx 35\omega$ we see no peak in the scattered energy, indicating the non-resonant operation. However, around $\omega_m=15.7\omega$, we see that the added resonance gives rise to a sharp peak in scattered power in the 1st harmonic, as it significantly increases the conversion efficiency.

%the Green function behaves like \\ $\lim_{z\rightarrow\infty} H_0^{(2)}(z)=\frac{e^{-jz}}{\sqrt{z}}$, so the currents and the waves act similar to plane waves. Hence the phase difference between the two dimer elements is $\Delta\phi\approx kd\cos(\theta)$.\\
%At certain frequencies $\omega_n=\omega_0+n\omega_m$, the corresponding wave-number $k_n=\frac{\omega_n}{C}$ (where $C$ is the speed of light) has certain values such that $\Delta\phi\approx kd\cos(\theta)\approx\pi$, meaning the currents will be different, hence enhancing the anti-symmetric mode. This means the sensitivity will get much better.\\
%On the other hand,we can also find values of $\omega_n, k_n$ such that $\Delta\phi\approx kd\cos(\theta)\approx2\pi$. In this case, the symmetric mode will be bigger, thus lowering the sensitivity. \\
%This is the effect that can be seen in fig. \ref{fig:contourLextended}.\\
%Notice that unlike the previous two effects used to enhance the sensitivity, this one has a different mechanism:\\
%While the frequency conversion and the parametric amplifier use resonance to enhance the current on each wire differently, the fast modulation effect uses the system innate architecture to force a phase difference, which affects the sensitivity. \\
%This can also be seen in fig. \ref{fig:contourLextended}(bottom) - looking at the ripples above $\omega_m> 20\omega_0$, we see no peaks in the energy of the system, due to the fact that the resonances are not involved.
\subsection{Parametric amplification}
Up to now, we have shown several ways the sensitivity benefits from incorporating time modulation. These revolved around the careful design of the dimer resonances, combined with leveraging the frequency conversion processes that occur when applying periodic temporal modulation. 
In circuits, parametric amplifiers amplify the input voltage by modulating one of the reactive system elements. The most simple case, which we will examine here, is the degenerate regime, where $\omega_{m}=2\omega_{0}$ \cite{collin1992foundations}. Recently, operating in this regime was also shown to enhance small antenna matching performance, and Q-factor \cite{AndreaChu,ChuParametric,ParametricChuAluExperiment}.
%, and phase between the modulation and the voltage: $\delta\phi=0$ will achieve maximum amplification, while $\delta\phi=\frac{\pi}{2}$ will reduce the output voltage amplitude.
Since the gain in this regime depends on the phase between the incoming signal and the modulation signal, we intuitively expect that this operation mode will both amplify the currents, and enhance the differences between as a function of the DoA, due to differential gain. 
%This approach may increase the sensitivity by first amplifying the currents, a
%Comparing our dimer system, by taking $\omega_m=2\omega_0$, we can have the same amplification in the dimer currents using the parametric amplifier principle. We also expect to see big effects on the sensitivity $S$, for two reasons:\\
%First, both currents will be amplified because of the parametric amplifier effect.\\
%In addition, we expect the difference between $I_1$ and $I_2$ to get bigger. That is because there is a phase difference between the two wires. Since the amplification depends on phase difference, each current will be amplified differently, thus enhancing the difference between the currents and so enhancing the sensitivity $S$.
%Perhaps show an analytical expression for the currents in this regime (assuming only two harmonics).
We start from Eq. \ref{eq:gammaMatrix}, \ref{eq:gMatrix} and \ref{eq:dimerMod}. We substitute $\omega_m=2\omega_0$, and assume, in general, a phase difference of $\delta\phi$ between the modulation and incoming signals. The incident fields, as they shuold be substituted into Eq. \ref{eq:gammaMatrix} are 
% \begin{equation}
%     [\Tilde{E}_1]=\begin{bmatrix}
%         \Tilde{E}^1_{-1} e^{-j\delta\phi}\\ \Tilde{E}^1_{0}e^{j\delta\phi}
%     \end{bmatrix}, \quad
%     [\Tilde{E}_2]=\begin{bmatrix}
%         \Tilde{E}^1_{-1}e^{-j\delta\phi-jkd\cos(\theta)} \\ \Tilde{E}^1_{0}e^{j\delta\phi+jkd\cos(\theta)}
%     \end{bmatrix}
% \end{equation}
\begin{equation}
\begin{split}
    [\Tilde{E}_1]&=\left[\hdots,0,\Tilde{E}_0^*e^{-j\delta\phi},\Tilde{E}_0e^{j\delta\phi},0,\hdots\right]^T \\
    [\Tilde{E}_2]&=
        \left[\hdots,0,\Tilde{E}_0^*e^{-j\delta\phi+jkd\cos\theta},\Tilde{E}_0e^{j\delta\phi-jkd\cos\theta},0,\hdots\right]^T.
    \end{split}
\end{equation}
The $\rtwo \Gamma,\rtwo G$ matrices would have the same structure as described in Eq. \ref{eq:gammaMatrix}. Using these to solve the currents in the wires results in current components at all frequencies $\omega=(1+2n)\omega_0$. However, since the main interaction we are interested in is the parametric amplification that results from coupling $[-\omega_0,\omega_0]$, we tune the system parameters differently now. We have already established that working at the resonance frequency of the anti-symmetric mode is preferable. In this case, it will have two key effects: first, the baseline sensitivity of the unmodulated dimer will be around the best we can obtain without modulation. Second, the parametric amplification will increase the content of this anti-symmetric mode. And lastly, operating near a system resonance will greatly simplify the analysis, rendering the currents for $\omega\neq\pm\omega_0$ negligible. Therefore, we use $C_0\approx 27.2pF$, and truncate $\rtwo \Gamma$ and $\rtwo G$ into $2\times 2$ matrices, for the main two frequencies in the system $[-\omega_0,\omega_0]$. In addition, we increase the losses to see the effects of parametric amplification more clearly, adding a $2\Omega$ resistance to the periodic loading.
\begin{equation}
\rtwo \Gamma_{2\times2}=
\begin{bmatrix}
         \gamma_0(-\omega_{0})+\frac{1}{-j\omega_{0}C_0\Delta} & \frac{m}{j\omega_{0}C_0\Delta} \\
        \frac{m^*}{-j\omega_{0}C_0\Delta} & \gamma_0(\omega_{0})+\frac{1}{j\omega_{0}C_0\Delta}   \\
    \end{bmatrix},\quad \rtwo G_{2\times2}=\begin{bmatrix}
    G(d,-\omega_0) & 0 \\
    0 & G(d, \omega_0)
    \end{bmatrix}
\end{equation}
Yielding the wire currents
\begin{equation}
    \begin{split}
        [\tilde{I}_2]&=\left[\rtwo \Gamma_{2\times 2} \rtwo G_{2\times 2}^{-1} \rtwo \Gamma_{2\times 2}- \rtwo G_{2\times 2}\right]^{-1}([\tilde{E}_2]+\rtwo \Gamma_{2\times 2} \rtwo G_{2\times 2}^{-1} [\tilde{E}_1])\\
        [\tilde{I}_1]&=\rtwo \Gamma_{2\times 2}^{-1}\left([\tilde{E}_1]+\rtwo G_{2\times 2}[\tilde{I}_2]\right)
    \end{split}
\end{equation}
%In the dimer, we have two resonance frequencies for the non-modulated case: the symmetric and the anti-symmetric. Hence, we expect only those 2 frequencies to contribute most of the energy, and we can look only at them. So, using the above equations, and using only $e^{\pm j\omega_0 t}$, we get:

%Notice that as stated earlier, phase is an important factor in parametric amplifiers: in the plates-capacitance model, we get maximum at $\delta\phi=0$ and minimum at $\delta\phi=\frac{\pi}{2}$. But in the model stated here, we get maximum at $\delta\phi=\frac{\pi}{2}$ and minimum at $\delta\phi=0$. This difference is due to the fact we measure the current on an inductance, which adds a factor of $\frac{\pi}{2}$ to the phase. If we divide our equation by $e^{j\frac{\pi}{2}}$, we will get the same points of maximum or minimum as the plates-capacitance model.
%Plot of S vs $\theta$ and $m$ (contour). 
\begin{figure}[H]
\centering
\includegraphics[width=11cm]{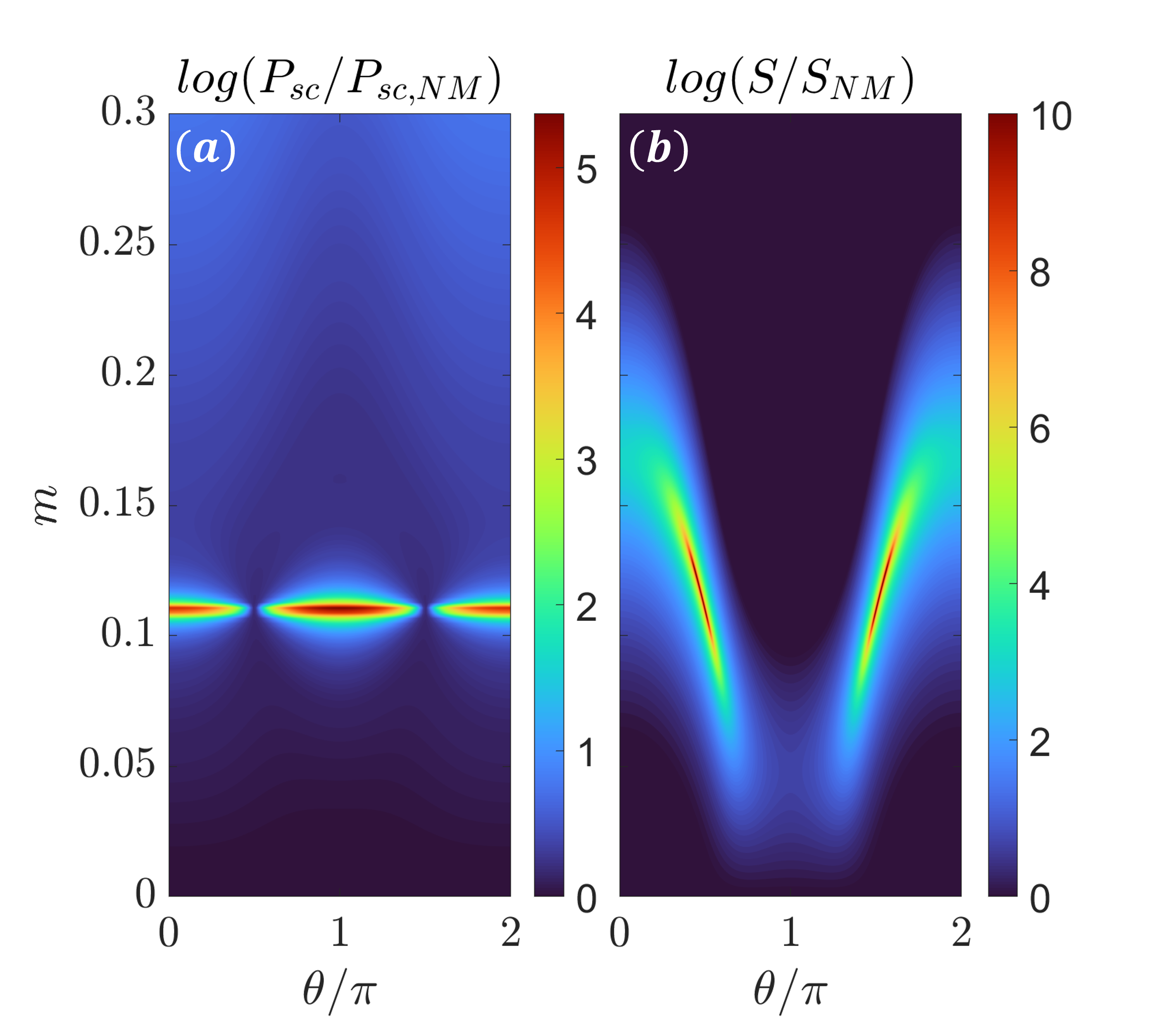}
\caption{(a) Total scattered power in the PA regime, normalized by the total scattered power of the non-modulated case. Strong amplification occurs only around $m\approx 0.11$. (b) $S(\theta)$, normalized by the sensitivity of the non-modulation case.}
\label{fig:SofthetaphasenonbestCparamamp}
\end{figure}
In fig. \ref{fig:SofthetaphasenonbestCparamamp}(a), the scattered power, normalized by the scattered power of the non-modulated case, is presented as a function of the modulation depth $m$ and of the DoA $\theta$, for $\delta\phi=0$. While there is a mild amplification overall, a very strong gain is obtained for a specific value $m\approx0.11$. This operation regime strongly resembles the negative impedance parametric amplifier \cite{collin1992foundations}, so such a strong response around a specific $m$ is expected. Working with this value of $m$ comes along with a sensitive dependence on the choice of other system parameters. However, since the mere amplification of currents is not the sole purpose, we now examine the map of $\mathcal S$. In Fig. \ref{fig:SofthetaphasenonbestCparamamp}(b), The normalized sensitivity is presented as a function of $m,\theta$. Here, we see that increased sensitivity can be obtained for a wide range of $m$ values. Moreover, the modulation depth can be used to tune the angle around which the optimal sensitivity is obtained, thus adding a great degree of flexibility that can be controlled by simple means. This shows that while the fundamental operation of the time-modulated system suggested here is similar to a parametric amplifier, the fact that other metrics play a key role is crucial, making this scheme relevant for a much broader range of parameters. 
\section{Conclusions}

%Explain how we can combine several approaches to leverage all the benefits. 
In this work, we have focused on studying the physical mechanisms that enhance a deeply-subwavelength dimer's sensitivity to the DoA angle. We have shown that by carefully tuning the modulation parameters, different frequency conversion processes contribute to increased sensitivity. Since the modulation parameters can be tuned, one obtains a highly flexible detector that can be tailored for wide-band operation and enhanced sensing in a specific, varying region of space. Modulating in a parametric amplification regime also contributes to a significant increase in sensitivity and exhibits tunability via the modulation depth and phase. 

When designing a DoA detection system, the different effects can be combined. As an example, one can stack two modulation tones, one to convert the incoming signal close to the desired anti-symmetric resonance and one to enhance it parametrically. 

Overall, we have shown that the richer physical interactions within the sensing system caused by the temporal modulation can be used to enhance DoA detection in various ways. This study can benefit many applications where the space occupied by the detection system must be extremely small, and many different branches can benefit from this research route.
% By using, for example, $\omega_m=2\omega_0$, the parametric amplifier effect will enhance the sensitivity. But by designing the system parameter such that the intersection in the resonance map between $n=1$ line and $\omega_{0,anti-resonance}$ line will be at $\frac{\omega_m}{\omega_0}=2$, the frequency translation effect can be stacked into a major sensitivity peak in the said area. \\
% Another way is combining the frequency conversion and the fast modulation: by designing the system such that the intersection between $\omega_1$ and $\omega_{anti-symmetric}$ to be at a high enough frequency, the fast modulation effect also amplifies the sensitivity: in fig. \ref{fig:contourLextended}, by adding an LC circuit to the periodic load, the peak sensitivity is located at $\omega_m\approx 15\omega_0$, where the fast modulation is also in effect. Thus, the sensitivity $S^1_{max}$ is much higher at this modulation frequency.   

\section{Acknowledgements}
Y. Mazor acknowledges support from the Israeli Science Foundation grant 1089/22.

\section*{Appendix: Incident and scattered power}\label{appendixA}
Let us assume that for a specific frequency we have the currents $I_1,I_2$ in the dimer wires. 
% Using Poynting Theorem, the energy of the non-modulated system is composed of:
% \begin{equation} \label{eq:PoyntingTheorem}
%     P_{sc}=\frac{1}{2}\Re \left\{ \iint (\rone{E}_{sc} \times \rone{H}_{sc}^*) \cdot \rone{dS} \right\}=\frac{1}{2}\Re \left\{ \iiint \rone{E}_{inc}\cdot\rone{J}^* dV \right\}=P_{inc}
% \end{equation}
% First, the calculation of $P_{sc}$. Start by calculating $E_{sc}$:
% \begin{equation}
% \begin{split}
%     \rone{E_{sc}}&=\left[G\left(\frac{d}{2}\right) I_1-\gamma|_{\frac{d}{2}}I_2\right] \hat{z}\\
%     &=\left[-\frac{\eta k}{4}H_0^{2}\left(\sqrt{\left(x-\frac{d}{2}\right)^2+y^2}\right)I_1-\frac{\eta k}{4}H_0^{2}\left(\sqrt{\left(x+\frac{d}{2}\right)^2+y^2}\right)I_2\right] \hat{z}
% \end{split}
% \end{equation}
% The medium is assumed to be lossless, we can integrate over $\sqrt{x^2+y^2}=\rho\rightarrow\infty$. Using Taylor's expansion, we can approximate 
% \begin{equation}\label{eq:taylorexpansion}
%     \begin{split}
%        k\sqrt{\left(x\pm\frac{d}{2}\right)^2+y^2}&=k\sqrt{x^2\pm dx+\frac{d^2}{4}+y^2}\quad\stackrel{x^2+y^2=\rho}{=}\quad k\rho\sqrt{1\pm\frac{dx}{\rho^2}+\frac{d^2}{4\rho^2}}=\\
%        &\stackrel{\frac{d}{\rho}\rightarrow0}{=}\quad k\rho(1\pm\frac{dx}{2\rho^2})=k\rho\pm\frac{kdx}{\rho}\quad\stackrel{\frac{x}{\rho}=\cos(\varphi)}{=}\quad k\rho\pm\frac{kd\cos(\varphi)}{2}
%     \end{split}
% \end{equation}
The far fields generated by the dimer are \cite{harrington2001time}
\begin{equation}
    \begin{split}
        \rone{E_{sc}}&= -\frac{\eta k}{4}\sqrt{\frac{2}{\pi k \rho}}\left[ I_1 e^{-jk\rho-j\frac{kd\cos(\varphi)}{2}-j\frac{\pi}{4}}+ I_2 e^{-jk\rho+j\frac{kd\cos(\varphi)}{2}-j\frac{\pi}{4}} \right]\hat{z}=\\
        &= -\frac{\eta k }{4}\sqrt{\frac{2}{\pi k \rho}}e^{-jk\rho-j\frac{\pi}{4}}\left[ I_1 e^{-j\frac{kd\cos(\varphi)}{2}}+ I_2 e^{j\frac{kd\cos(\varphi)}{2}} \right]\hat{z}
    \end{split}
\end{equation}
and the corresponding magnetic field is 
\begin{equation}
    \rone{H_{sc}}=\frac{1}{\eta}\hat{\rho}\times\rone{E_{sc}}=\frac{1}{\eta} E_{sc}\ \hat{\varphi}.
\end{equation}
This yields the scattered power per unit length
\begin{equation}
\begin{split}
    P_{sc}&=\frac{1}{2}\Re\int_0^{2\pi}\boldsymbol{E}_{sc}\times\boldsymbol{H}_{sc}^*\rho d\varphi \\
    &=\frac{\eta^2 k^2 \rho }{16}\frac{2}{\pi k \rho}|e^{-jk\rho-j\frac{\pi}{4}}|^2 \int_0^{2\pi} d\varphi \left| I_1 e^{-j\frac{kd\cos(\varphi)}{2}}+ I_2 e^{j\frac{kd\cos(\varphi)}{2}} \right|^2    
\end{split}
\end{equation}
Performing the integration, we get
\begin{equation}
    \begin{split}
        P_{sc}&= \frac{\eta^2 k^2 \rho }{16}\frac{2}{\pi k \rho}|e^{-jk\rho-j\frac{\pi}{4}}|^2 \int_0^{2\pi} d\varphi (\left| I_1 \right|^2+ \left| I_2 \right|^2)+ 2\Re\left\{|I1|\cdot |I_2| e^{-jkd\cos(\varphi)-\angle(I_1,I_2)}\right\}\\
        &=\frac{\eta^2 k}{4}\left(\left[|I_1|^2+|I_2|^2\right]+2J_0(kd)\Re\left\{I_1\cdot I_2^*\right\}\right)
    \end{split}
\end{equation}
Next, $P_{inc}$ is composed of
\begin{equation}
     P_{inc}=\frac{1}{2}\Re\left\{E^{inc}_{wire}I^*_{wire}\right\}+\frac{1}{2}\Re\left\{E^{inc}_{other-wire}I^*_{other-wire}\right\}
\end{equation}
For each wire, the current can be expressed as:
\begin{equation}
    I_{wire}=\alpha E_{wire}=\alpha\left(E^{inc}+G(d)I_{other-wire}\right)
\end{equation}
And since the problem is symmetric, we can use it for each wire. Hence:
\begin{equation}
    P_{inc}=\frac{1}{2}\Re\left\{\alpha^*(|E^{inc}_1|^2+E^{inc}_1 G^*(d)I^*_{2})\right\}+\frac{1}{2}\Re\left\{\alpha^*(|E^{inc}_2|^2+E^{inc}_2 G^*(d)I^*_{1})\right\}
\end{equation}
When modulation is present, we need, in general, to sum up all the contribution from all existing harmonics.
% In order to add time modulation to the calculated quantities, we replace each quantity with its vector or matrix counterpart:
% \begin{equation}
%     \begin{split}
%         \alpha &\stackrel{eq.\ref{eq:gammaMatrix}}{\longrightarrow}\underline{\underline{\alpha}}\\
%         G &\stackrel{eq.\ref{eq:gMatrix}}{\longrightarrow} \underline{\underline{G}}\\
%         I &\stackrel{eq.\ref{eq:gammaMatrix}}{\longrightarrow} [I]\\
%         E &\stackrel{eq.\ref{eq:gammaMatrix}}{\longrightarrow} [E]\\         
%     \end{split}
% \end{equation}

% \clearpage
\bibliographystyle{ieeetr}
% %\bibliographystyle{abbrv}
\bibliography{Paper1}
% \begin{thebibliography}{9}
% \bibitem[Doe]{doe} \emph{First and last \LaTeX{} example.},
% John Doe 50 B.C. 
% \end{thebibliography}

\end{document}